\documentclass[twocolumn, aps, nofootinbib, letterpaper]{revtex4}

\usepackage{amsmath,amssymb}
\usepackage{graphicx}
\usepackage{color}

\newcommand{\Lie}[0]{{\cal L}\, }

\newcommand{\tq}{\tilde{q}}

\newcommand{\bea}{\begin{eqnarray}}
\newcommand{\eea}{\end{eqnarray}}

\begin{document}

\title{Extremality conditions for isolated and dynamical horizons}
\author {Ivan Booth\footnote{E-mail:
ibooth@math.mun.ca}}
\affiliation{
Department of Mathematics and Statistics, Memorial University of
Newfoundland \\ 
St. John's, Newfoundland and Labrador, A1C 5S7, Canada}
\author{Stephen Fairhurst \footnote{E-mail: Stephen.Fairhurst@astro.cf.ac.uk}}
\affiliation{Department of Physics, University of Wisconsin--Milwaukee
Milwaukee, Wisconsin, 53201, USA \\
LIGO - California Institute of Technology,
Pasadena, CA  91125, USA\\
School of Physics and Astronomy, Cardiff University,
Cardiff, CF2 3YB, United Kingdom}

\begin{abstract}

A maximally rotating Kerr black hole is said to be extremal. In this
paper we introduce the corresponding restrictions for isolated and
dynamical horizons. These reduce to the standard  notions for Kerr but
in general do not require the horizon to be either stationary or
rotationally symmetric. We consider physical implications and 
applications of these results. In particular we introduce a parameter
$e$ which characterizes how close a horizon is to extremality and 
should be calculable in numerical simulations. 

\end{abstract}

\maketitle

\section{Introduction}
\label{sec:intro}

It is well known that there is a limit on the maximum allowed angular
momentum for a Kerr black hole. If such a hole has mass $M$ then the
angular momentum $J$ must satisfy $J \leq M^2$. Solutions which saturate
this bound are known as extremal while Kerr spacetimes with $J>M^2$
contain naked singularities rather than black holes. Given this
constraint on stationary solutions it is natural to consider whether
there is a similar restriction for astrophysical black holes. In
contrast to the Kerr holes which sit alone in an otherwise empty
universe, real black holes do not exist solely in isolation and can, for
example, be surrounded by accretion disks or be components of binary
systems. 

For this reason, it is interesting to investigate whether extremality
conditions can be formulated and applied to interacting black holes.
This is of particular interest during black hole collisions.  It is
widely accepted that following a merger, the final black hole will
settle down to one of the known stationary solutions.  However, during
the highly dynamical merger phase, it is not clear whether the black
hole's angular momentum is bounded.  The existence or lack of an
extremality condition may help us to understand the physics of black
hole mergers, and provide insight into whether binary black holes will
necessarily ``hang up'' in orbit, emitting excess angular momentum prior
to forming a common horizon. Similar questions arise for black holes
forming from the gravitational collapse of matter. 

Away from the Kerr--Newmann family of solutions there are some recent
results that either support or cast doubt on the possible existence of
such a bound.  In support, Dain \cite{Dain:2006wb,
Dain:2005vt,Dain:2005qj} has shown that for a large class of
asymptotically flat, axially symmetric, vacuum black holes $J_{ADM} \leq
M^2_{ADM}$ where the subscripts indicate that these are the ADM mass and
angular momentum as measured at spatial infinity. By contrast Petroff
and Ansorg \cite{Ansorg:2005bq, Ansorg:2006qt, Petroff:2005vu} have
recently generated numerical examples of black holes surrounded by
rotating rings of matter for which the Komar mass and angular momentum
violate the bound $J_{\mathrm{Komar}} \leq M^{2}_{\mathrm{Komar}}$.

Clearly in considering these issues one needs to be careful about how
the physical quantities are defined. In particular, in formulating a
bound one would like to distinguish between the mass and angular
momentum directly associated with the black hole  versus any matter or
gravitational waves surrounding it. This is, of course, easier said than
done. Mass and energy are notoriously ambiguous quantities in general
relativity.  They are well-defined for entire asymptotically flat
spacetimes but in general it is not possible to assign mass and energy
to more localized regions of spacetime (see, for example, the discussion
in \cite{Szabados:2004vb}). Similar problems arise for angular momentum
and away from axi-symmetry it is not at all clear that angular momentum
can be described by a single number. 

Distinguishing between local and global properties of black holes is one
of the main motivations for the recent interest in quasilocal
characterizations of horizons, including trapping  \cite{Hayward:1993wb,
Hayward:2004fz, Hayward:2006he, Hayward:2004dv}, isolated
\cite{Ashtekar:1998sp, Ashtekar:1999yj, Ashtekar:2000hw,
Ashtekar:2000sz,Ashtekar:2001is}, and dynamical \cite{Ashtekar:2002ag,
Ashtekar:2003hk} horizons. In this paper we apply the machinery
developed in the study of quasilocal horizons to investigate local
characterizations of extremality.

We will argue that the ambiguities in defining mass and angular momentum
mean that, in general, the usual Kerr extremality bound is not well
formulated for quasilocal horizons. Thus we will examine alternative
characterizations of extremality and show that they do apply. For
isolated horizons these arise from 1) the non-negativity of the surface
gravity and 2) the idea that there should be trapped surfaces just
inside the horizon.  For dynamical horizons only the second these
characterizations is applicable since the surface gravity is only
meaningful for slowly evolving dynamical horizons (in the sense of
\cite{Booth:2003ji,Kavanagh:2006qe,Booth:2006bn}). 

Both the surface gravity and trapped surface characterizations of
extremality give rise to an alternative, local extremality condition.
This condition is similar in spirit to the Kerr extremality bound, and
involves contributions from both the black hole's angular momentum and
horizon matter fields.  However, it can be written locally on the
horizon and angular momentum ambiguities are avoided by making use of
the square of an ``angular momentum density'' integrated over the
horizon rather than the angular momentum relative to any particular
axis.  With this definition of extremality, we show that generic
dynamical horizons, in the sense of \cite{Ashtekar:2005ez}, are
necessarily sub-extremal.  Isolated horizons can obtain extremality (in
which case the local geometry must be that of the Kerr horizon
\cite{Lewandowski:2002ua}).  We discuss how this local extremality
compares to the standard Kerr relation, and argue that in those cases
where both are well formulated it may still be possible for a black hole
to violate the Kerr bound.
 
The paper is laid out as follows.  We begin in Section
\ref{sec:stationary} with a discussion of extremality for stationary
black holes and recall the three notions of extremality already
mentioned: maximum angular momentum, vanishing surface gravity, and
coincidence of the inner and outer horizons. The next section shows how
these notions may be adapted to isolated horizons, examines conditions
under which they are equivalent, and considers situations under which
one or more of them might be violated.  In Section \ref{sec:dynamical}
we use this experience to study the equivalent notions for dynamical
horizons and show that these horizons are always sub-extremal. Finally,
Section \ref{sec:summary} provides a brief summary.  An Appendix shows
how the various notions apply to the Kerr (anti-)deSitter family of
solutions. 

\section{Stationary Black Hole Horizons}
\label{sec:stationary}

Let us begin by reviewing the notion of extremality for stationary,
asymptotically flat black hole space-times.  In Einstein--Maxwell
theory, the uniqueness theorems tell us that the class of such solutions
is restricted to the Kerr--Newman space-times.  Furthermore, these
space-times are characterized by only three quantities; their mass M,
angular momentum J and electric charge Q.  Black hole solutions exist
for all values of $M$, $J$ and $Q$ which satisfy the inequality:
\begin{equation}\label{eq:kerr_bound}
  a^2 + Q^{2} \le M^{2} \quad \mathrm{where} \quad a = J/M\, .
\end{equation}
If this inequality is violated the resulting spacetimes are still
solutions of the Einstein equations, however they contain a naked
singularity in lieu of a black hole.  The first notion of
extremality arises from Eq.~(\ref{eq:kerr_bound}).  Solutions for which
the equality is satisfied, namely
\begin{equation}\label{eq:kerr_extremality}
  a^2 + Q^2 = M^2
\end{equation}
are said to be extremal as they contain the maximum allowed
angular momentum/charge for a given mass. 

Since the Kerr--Newman solutions are stationary and axi-symmetric, they
have both a time-translation Killing vector field $t^{a}$ and a
rotational Killing vector field $\phi^{a}$.  The event horizon is a
non-expanding null surface whose null normal is
\begin{equation}\label{eq:kerr_xi}
  \xi^{a} = t^{a} + \Omega \phi^{a} \, ,
\end{equation}
where $\Omega$ is interpreted as the angular velocity of the horizon.
It is then straightforward to calculate the acceleration of $\xi$ at the
horizon:
\begin{equation}\label{eq:surf_grav}
  \xi^b \nabla_{b} \xi^{a} = \kappa \xi^{a} \, .
\end{equation}
The quantity $\kappa$ is known as the surface gravity.  Since $\xi$ is
defined in terms of Killing vectors which are appropriately normalized
at infinity, there is no ambiguity in its normalization and by direct
calculation
\begin{equation}\label{eq:kerr_surf_grav}
  \kappa = \frac{ \left( M^{2} - a^{2} - Q^{2} \right)^{1/2} }
    {2M \left[ M + \left( M^{2} - a^{2} - Q^{2} \right)^{1/2} \right]
      - Q^2} \, .
\end{equation}

The second notion of extremality comes from this surface gravity.  It is
clear from Eqs.~(\ref{eq:kerr_bound}) and (\ref{eq:kerr_surf_grav})
that the surface gravity is only well-defined for black hole (as opposed
to naked singularity) solutions
and is necessarily non-negative.  Furthermore, for
an extremal black hole satisfying (\ref{eq:kerr_extremality}), the
surface gravity vanishes, i.e. $\kappa = 0$.  Thus, vanishing surface
gravity is often taken as the defining property of an extremal horizon.  

Finally, we can understand extremality from the geometric structure
of space-time 
--- one of the fundamental properties of a black hole is that
it contains \emph{trapped surfaces} which are defined in the following
way.  Any spacelike two-surface has two future-pointing null normals,
which we will denote $\ell$ and $n$.  Then, the expansion of these null
vectors is defined as:
\begin{equation}\label{eq:expansions} 
  \theta_{(\ell)} = \tq^{ab} \nabla_{a} \ell_{b} 
  \quad \mbox{ and } \quad 
  \theta_{(n)} = \tq^{ab} \nabla_{a} n_{b} \, , 
\end{equation} 
where $\tq_{ab}$ is the metric of the two-surface.  On a trapped
surface, the expansions of both null vector fields are negative.  This
is in contrast to a typical (convex) two-surface in flat space which
will have one positive and one negative expansion.  For asymptotically
flat spacetimes, the existence of a trapped surface is sufficient to
imply the existence both an event horizon enclosing the surface and a
space-time singularity somewhere in its interior \cite{Penrose:1964wq}. 

For typical charged or rotating black holes, there are two horizons, the
event horizon and the inner Cauchy horizon. These are null, foliated by
two-dimensional marginally trapped surfaces ($\theta_{(\ell)} = 0$ and
$\theta_{(n)} < 0$), and split the space-time into distinct regions. It
is only in the region between the horizons that trapped surfaces exist.
If the charge or angular momentum is increased towards the extremal
value, the trapped region between the horizons shrinks, until, at
extremality, the inner and outer horizons coincide, the trapped region
vanishes and only the marginally trapped surfaces of the horizon remain.
It is this notion of extremality which is used in Israel's proof that a
non-extremal black hole cannot achieve extremality in a finite time
\cite{PhysRevLett.57.397}.

Thus, we see that for event horizons in stationary space-times, there
are three notions of horizon extremality which all coincide:

\begin{description}

\item[First Characterization:] The angular momentum and charge of a
black hole are restricted according to Eq.~(\ref{eq:kerr_bound}).  For
an extremal black hole, $a^2 + Q^2 = M^2$. 

\item[Second Characterization] The surface gravity $\kappa$ of a black
hole must be greater than or equal to zero.  The surface gravity
vanishes if and only if the horizon is extremal.

\item[Third Characterization] The horizon of the black hole is a
marginally trapped surface.  For non-extremal black holes, the interior
of the black hole must contain trapped surfaces, while for extremal
black holes, the inner and outer horizons coincide and there are no
trapped surfaces. 

\end{description}

In the remainder of this paper, we will argue that the second and third
definitions can be extended to isolated and dynamical horizons.
Furthermore, we will obtain a horizon relation similar in spirit, though
not identical, to the one appearing in the first definition above.  We
start with isolated horizons. 

\section{Isolated Horizons}

Isolated horizons have been introduced to capture the local physics of
the horizon of a black hole in equilibrium \cite{Ashtekar:1998sp,
Ashtekar:1999yj, Ashtekar:2000hw, Ashtekar:2001is, Ashtekar:2000sz}.
These are null surfaces and so form causal boundaries.  However, unlike
event horizons, they are defined (quasi-)locally. Specifically, a null
surface $\Delta$ of topology $S^2 \times \mathbb{R}$ with (degenerate)
metric $q_{ab}$, derivative $D_a$, and normal $\ell_a$ is an isolated
horizon if: 

\begin{enumerate}
\item $\Delta$ is non-expanding: $\theta_{(\ell)} = 0$ , 

\item an energy condition holds at the horizon: $-T^a_{\; \; b} \ell^b$
is future-directed and causal, and

\item the null vector $\ell^a$ is scaled such that
\begin{equation}\label{eq:ih_extrinsic}
  [\Lie_{\ell}, D] = 0 \, .
\end{equation}

\end{enumerate}

The energy condition is weaker than and implied by any of the standard
energy conditions.  Together with the first condition and the
Raychaudhuri equation it follows both that the intrinsic geometry of $\Delta$
is invariant in time: $\mathcal{L}_\ell q_{ab} = 0$ and that there is
no flux of matter through the horizon: $T_{ab} \ell^a \ell^b = 0$. The
third condition  fixes the scaling of $\ell^a$ up to an overall
constant and ensures that the extrinsic geometry is similarly invariant
in time. 

To make all of this a little more concrete, note that one can always
find functions $v$ on $\Delta$ that are compatible with $\ell$ (so that
$\mathcal{L}_\ell v = 1$) and which have spacelike level surfaces $S_v$
with topology $S^2$. For such a function, $n_a = - D_a v$ is not only
normal to these surfaces of constant $v$ but is also null and satisfies
$\ell \cdot n = -1$.  The spacelike metric on the $S_v$ can be written
as $\tilde{q}_{ab} = g_{ab} + \ell_a n_b + n_a \ell_b$. 

With these additional structures the invariance of the intrinsic
geometry can be written as
\bea
\mathcal{L}_\ell \tilde{q}_{ab} = \frac{1}{2} \theta_{(\ell)} \tilde{q}_{ab} + 
\sigma^{(\ell)}_{ab} = 0 \, , 
\eea
so that both the expansion $\theta_{(\ell)}$ and shear
$\sigma^{(\ell)}_{ab}$ of the two-surfaces vanish. Further, the third
condition implies that the corresponding expansion $\theta_{(n)}$ and
shear $\sigma^{(n)}_{ab}$ in the $n$-direction are also invariant in
time
\bea
\mathcal{L}_\ell \theta_{(n)} = 0 \; \; \mbox{and} 
\; \; \mathcal{L}_{\ell} \sigma^{(n)}_{ab} = 0 \, ,
\eea
as is the connection 
\bea
\tilde{\omega}_a = - \tilde{q}_a^b n_c \nabla_b \ell^c
\label{eq:omega}
\eea 
on the normal bundle to the foliation two-surfaces:
\bea
\mathcal{L}_\ell \tilde{\omega}_a = 0 \, . 
\eea
Finally, one can use the axioms to prove a zeroth law. For the allowed 
scalings of the null vectors, the surface gravity $\kappa$, defined in a 
similar manner to that on the event horizon (\ref{eq:surf_grav}), 
\bea
\label{eq:iso_surf_grav}
\ell^b \nabla_b \ell^a \equiv \kappa \ell^a \, 
\eea
is constant on the horizon: $D_a \kappa = 0$ \cite{Ashtekar:2000hw}.

On an isolated horizon, the scaling of the null vectors is only fixed up
to an overall positive multiplicative constant. Under allowed rescalings
$\ell \rightarrow c \ell$ and $n \rightarrow n/c$, the connection
$\tilde{\omega}_a$ is invariant while $\kappa \rightarrow c \kappa$.
Thus, while $\kappa$ is constant over $\Delta$, its exact value is
only fixed up to sign (ie.\ positive, negative, or zero). 

Derivations of these facts can be found in the already cited references
or in \cite{Ashtekar:2001jb} which focuses on the geometry of horizons.

\subsection{Extremality from $Q$, $a$ and $M$?}

We begin with the first notion of extremality: the horizon of a
Kerr--Newman black hole is extremal if and only if $a^2 + Q^2 = M^2$.
Let us attempt to extend this to isolated horizons.  The prerequisite to
this is to obtain a satisfactory definition of each of these quantities
and, except for the electric charge, that is where problems arise.
Given a two-dimensional cross-section $S_v$ of the horizon, the charge
is well-defined by Gauss' law:
 \bea
Q \equiv \frac{1}{4 \pi} \int_{S_v} \mspace{-10mu} d^2 x 
  \sqrt{\tilde{q}} F_{ab} \ell^a n^b 
= \frac{1}{4 \pi} \int_{S_v} \mspace{-10mu} d^2 x
   \sqrt{\tilde{q}} E_{\perp} \, ,
\label{Charge}
\eea
where $\tilde{q}$ is the determinant of the two-metric $\tilde{q}_{ab}$
on $S_v$ and $F_{ab}$ is the electromagnetic field tensor. Equivalently,
we rewrite $F_{ab} \ell^a n^b  = F_{ab} \hat{u}^a \hat{s}^b =: E_{\perp}$
where $\hat{u}$ and $\hat{s}$ respectively are orthogonal timelike and
spacelike unit normal vectors to $S_v$. $E_{\perp}$ is the flux of the
electric field through $S_v$ as observed by a timelike observer with
evolution vector $\hat{u}^a$. Since the horizon is isolated, this is
independent of the cross-section \cite{Ashtekar:2000hw}. 

Next, we consider angular momentum. In classical, non-relativistic,
physics angular momentum is defined relative to an axis of rotation. For
isolated horizons the analogue of an axis of rotation is a rotational
vector field. Following \cite{Booth:2003ji,Booth:2005ss,Booth:2006bn}
this is given by $\phi^a \in TS_v$ whose flow foliates the $S_v$ into
closed integral curves of parameter length $2 \pi$ plus two fixed points
(the poles of the rotation).  A vector field of this type is necessarily
divergence-free and the canonical example is a horizon with a rotational
Killing vector field $\phi^a$ so that 
\bea
\mathcal{L}_\phi \tq_{ab} = 0 \, . 
\eea

The angular momentum relative to a rotational vector field $\phi^a$ is then 
\cite{Ashtekar:2001is, Ashtekar:2000sz}
\begin{equation}\label{eq:angular_momentum}
  J[\phi] = \frac{1}{8\pi G} \int_{S} 
  d^2 x  \sqrt{\tilde{q}} \phi^{a} \tilde{\omega}_{a} +
  \frac{1}{4\pi G} \int_{S} d^{2} x \sqrt{\tilde{q}} \phi^{a} A_{a} E_{\perp}
  \, ,
\end{equation}
where $\tilde{\omega}_a$ is the connection of the normal bundle that we
have already encountered in Eq.\ (\ref{eq:omega}), and $A_{a}$ is the
electromagnetic connection. It is immediate that on an isolated horizon
this quantity is independent of the choice of cross-section $S_v$. 

This is closely related to other standard measures of
angular momentum such as the Brown-York \cite{Brown:1992br} or dynamical horizon \cite{Ashtekar:2003hk} measures. 
In particular, as is discussed in more detail in
\cite{Ashtekar:2000hw}
\bea
D_{a} \ell^{b} = (- \kappa n_a + \tilde{\omega}_a) \ell^b 
\eea
is the Weingarten map and is analogous to the standard extrinsic
curvature, although tailored to the null surface of the horizon. Then,
it is not surprising that the geometric part of the angular momentum
(\ref{eq:angular_momentum}) agrees with usual extrinsic curvature
formula. To see this, consider the case where the isolated horizon is an apparent horizon found in a numerical simulation. In this case, the
$S_v$ are each contained in spacelike three-surfaces $\Sigma_t$,
$\hat{u}^a$ is the future directed unit normal to the $\Sigma_t$ and
$\hat{s}^a \in T \Sigma_t $ is the outward-pointing spacelike unit
normal to the $S_v$.  Then, for a divergence-free rotational vector
field $\phi^{a}$ it is straightforward that  the 
isolated horizon angular momentum can be rewritten as
\begin{equation}
  J[\phi] = \frac{1}{8\pi G} \int_{S_v} d^2 x \sqrt{\tilde{q}} \,
  \left( 
  \phi^{a} \, \hat{s}^{b} \, K_{ab} +
 2 \phi^{a} A_{a} E_{\perp}
  \right)
  \, ,
\label{eq:JNum}
\end{equation}
where $K_{ab} = h_a^c h_b^d \nabla_{c} \hat{u}_{d}$ is the extrinsic
curvature of $\Sigma_t$ (with $h_{ab}$ the induced three-metric on
$\Sigma_t$).  In stationary blak hole spacetimes, this measure also
agrees with the Komar and ADM angular momenta evaluated at infinity. 

Thus, given a rotational vector field $\phi^a$ the angular momentum of
the horizon is well-defined.  Unfortunately, in the absence of
axi-symmetry there is no obvious way to uniquely select a geometrically
preferred rotational vector field. Indeed, for highly distorted
horizons, it is by no means clear that one should always expect such an
``axis of rotation" to exist. For non-axisymmetric horizons it seems
unlikely that the angular momentum can be characterized by a single
number (though see \cite{Dreyer:2002mx, Schnetter:2006yt, Hayward:42ss,
Cook:2007wr, Korzynski:2007hu} for alternative viewpoints). 

Even if we restrict our attention to axially symmetric horizons, in
order to define extremality by Eq.~(\ref{eq:kerr_bound}) we would still
need a definition of horizon mass $M$ and here the greatest difficulties
arise. While local definitions of angular momentum for a surface are
readily available and tend to agree, the issue of a local energy or mass
is much more difficult \cite{Ashtekar:2000hw, Booth:2005ss,
Szabados:2004vb}.  In particular, the rescaling freedom of $\ell$
precludes the identification of a preferred energy associated with
evolution along $\ell$.  A common solution to this is to simply
\textit{define} the mass of a horizon with area $A$ and angular momentum
$J$ to be equal to the value it would take in the Kerr space-time.  Then
by the Christodoulou formula \cite{PhysRevLett.25.1596}
\begin{equation}\label{eq:kerr_mass}
  M^{2} := \frac{(R_H^{2}+Q^2)^2+4 J^2}{4 R_H^2} 
\end{equation}
where $R_H = \sqrt{a/(4 \pi)}$ is the areal radius of the horizon. With
this mass, it is straightforward to show that $|J|$ is less than $M^2$.
However this is simply a property of the definition and the physical
relevance of (\ref{eq:kerr_mass}) for highly distorted black holes is,
at best, unclear. 

Given the difficulties with the definition of mass, it probably makes
more sense to rephrase any characterization of type
(\ref{eq:kerr_bound}) entirely in terms of quantities such as $R_H$,
$Q$, and  $J$ which can be locally measured on the horizon. Then, this
bound may be rewritten as
\bea
\label{eq:new_kerr_bound}
Q^4 + 4 J^2  \leq R_H^4 \, . 
\eea
Thus defining $Q$, $J$, and $R_H$ as we have above, this is the first
possible characterization of extremality, at least for axi-symmetric
horizons.  There is no general derivation that this bound must hold
outside of the Kerr--Newman family.  Indeed, the inequality in
Eq.~(\ref{eq:new_kerr_bound}) can be violated for non-asymptotically
flat spacetimes (Appendix \ref{KadS}). Similarly, in higher dimensional 
asymptotically flat space-times the original bound
(\ref{eq:kerr_bound}) can also be violated \cite{Myers:1986un,
Emparan:2003sy}.  Despite this, Ansorg and Pfister have demonstrated
that (\ref{eq:new_kerr_bound}) holds (with equalilty) for a class of
extremal configurations of black holes surrounded by matter rings.
Furthermore, they have conjectured that the inequality will hold for
stationary, axially and equatorially symmetric black hole--matter ring
solutions \cite{Ansorg:2007fh}.

\subsection{Extremality from $\kappa$}
 \label{Sect:kappa}
 
The second notion of extremality for Kerr black holes says that the
surface gravity is positive for sub-extremal holes and zero for extremal
holes. It is never negative.  We now consider  this characterization for
isolated horizons and see that in many ways it is more satisfactory than
that considered in the previous section. 
 
The surface gravity for an isolated horizon was given in Eq.
(\ref{eq:iso_surf_grav}), where it was noted that a zeroth law holds so
that $\kappa = \kappa_o$ everywhere on $\Delta$ for some (fixed)
$\kappa_o \in \mathbb{R}$. The scaling of the null vectors is only fixed
up to a positive multiplicative  constant so rescalings of the form
$\ell \rightarrow c \ell$, $c \in \mathbb{R}^+$ are allowed. For such
rescalings $\kappa \rightarrow c \kappa$ and so the formalism only
allows us to say whether an isolated horizon has a surface gravity that
is positive, negative, or zero.  This is sufficient for our purposes; an
isolated horizon is sub-extremal if and only if $\kappa > 0$, extremal
if $\kappa = 0$, and super-extremal if $\kappa < 0$. 

Such a definition is more than just nomenclature. For $\kappa = 0$ there
is a local uniqueness theorem for isolated horizons; the intrinsic
geometry of an extremal isolated horizon must be identical to that of
the Kerr--Newmann horizon with the same area, charge and angular
momentum \cite{Lewandowski:2002ua}.  Further, sub-extremal horizons must
obey bounds on their electric charge and the angular momentum one-form
$\tilde{\omega}_a$.  To see this, note that the the evolution of
$\theta_{(n)}$ is given by \cite{Ashtekar:2000hw,Booth:2006bn}:
\begin{eqnarray}\label{eq:lie_l_theta_n}
  &&\Lie_{\ell} \theta_{(n)} 
    + \kappa \theta_{(n)} 
    + \tilde{R}/2 
    = \nonumber \\
 && \qquad \tilde{d}^{a} \tilde{\omega}_{a} 
    + \tilde{\omega}^{a} \tilde{\omega}_{a} 
    + (8 \pi G) T_{ab} \ell^{a} n^{b} \, ,
\end{eqnarray}
where $\tilde{R}$ is the two-curvature of the cross-sections of the
horizon, $\tilde{d}$ is the spacelike two-metric compatible covariant
derivative and $T_{ab}$ is the energy-momentum tensor. 

The standard definition of an isolated horizon doesn't restrict the sign
of the inward expansion $\theta_{(n)}$.  However, since we are
interested only in black hole horizons%
\footnote{The defining conditions for isolated horizons are intended to
be necessary conditions that a null surface should meet in order to be
considered as the boundary of a non-interacting black hole region.
However, they are not sufficient to distinguish black hole horizons from
white hole or cosmological horizons.  Furthermore, there are examples of
isolated horizons which either do not correspond to black holes (in
\cite{Pawlowski:2003ys} there are no trapped surfaces) or their black
hole status is unclear (in \cite{Fairhurst:2000xh} it is not known
whether the ``black holes" all contain trapped surfaces).}%
, it is reasonable to impose the extra requirement that there be trapped
surfaces ``just inside" the horizon. Thus, by continuity we restrict our
attention to horizons with $\theta_{(n)} < 0$. 

Now, consider equation (\ref{eq:lie_l_theta_n}) evaluated on a
sub-extremal isolated horizon.  On the left-hand side, the first term
will vanish since the geometry is time independent.  The second term is
necessarily non-positive: by assumption surface gravity is non-negative,
and we have restricted to black hole horizons where $\theta_{(n)} < 0$.
Finally, although the $\tilde{R}$ itself can vary in sign, we know that
since the cross-sections of the horizon have topology $S^2$, $\int_{S_v}
\sqrt{\tilde{q}} \tilde{R} = 8 \pi$.

Thus, integrating (\ref{eq:lie_l_theta_n}) over any $S_v$ gives
\begin{equation}\label{eq:extremality}
  e := \frac{1}{4\pi} \int_{S_v} d^2 x \sqrt{\tilde{q}} 
   \left( 8 \pi G T_{ab} \ell^{a} n^{b} + ||\tilde{\omega}||^{2} \right) 
  \le 1 \, ,
\end{equation}
(the integral of the exact derivative $\tilde{d}_{a} \tilde{\omega}^{a}$
vanishes). This gives an alternative characterization of extremality for
isolated horizons: $\kappa$ vanishes if and only if $e =1$ and is
positive if and only if $e < 1$.  This expression provides a local
extremality condition for isolated horizons expressed in terms of the
horizon angular momentum (encoded in the one-form $\tilde{\omega}$) and
matter fields.  As such, it is similar in spirit to the standard Kerr
bound.  However, this condition is applicable to all isolated horizons,
and does not require either axi-symmetry or asymptotic flatness.

The exact interpretation of the matter term depends on the matter
present at the horizon, but for electromagnetism it is related to the
electric and magnetic charge of the hole: 
\bea
8 \pi G T_{ab} \ell^a n^b = \left(E_{\perp}^2 + B_{\perp}^2  \right) \, . 
\eea
The first term is the square of the electric flux density, while the
second is the square of the corresponding magnetic flux --- the integral
of $B_\perp$ is the magnetic charge contained by $S_{v}$.   

The second term of (\ref{eq:extremality}) is associated with angular
momentum.  It side-steps the problems of defining an axis of rotation by
working with the square of the angular momentum density integrated over
the horizon as a measure of the total angular momentum.  To gain some
insight into this quantity, consider an axi-symmetric horizon.  On such
a horizon, the angular momentum can be decomposed into its multipole
moments \cite{Ashtekar:2004gp}.  Based on this decomposition, $J$ can be
interpreted as the dipole angular momentum of the horizon.  Then the
quantity appearing in our extremality condition (\ref{eq:extremality})
can, at least intuitively, be thought of as the sum of the squares of
\textit{all} the angular momentum multipoles of the horizon.

These associations can be made concrete by calculating $e$ for the known
black hole solutions. First for Reissner--Nordstr\"om space-times (where
$\tilde{\omega}_a$ vanishes) it is straightforward to show that 
\begin{equation}
 e = Q^{2}/R_H^{2}
\end{equation}
where $Q$ is the charge and $R_H$ is the areal radius of the horizon.
Thus $e = 1$ corresponds to the extremality condition $Q = R_H ( = M )$
while $e < 1$ implies that $Q < R_H$. 

\begin{figure}
\includegraphics{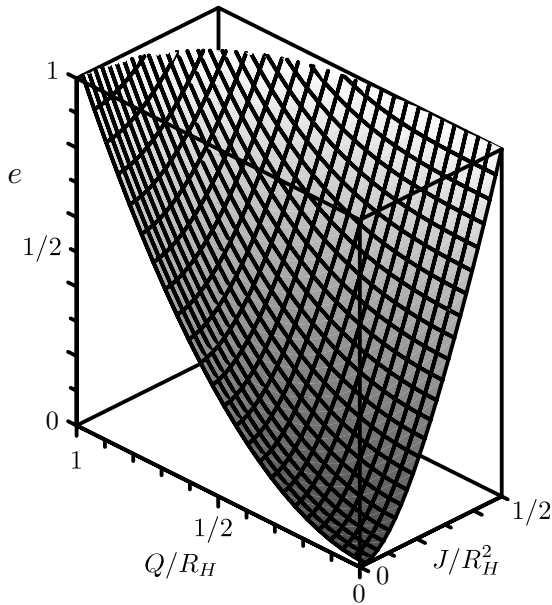}
\caption{Plot of the extremality parameter, $e$, for a Kerr--Newman
horizon as a function of the scaled angular momentum $J/R_H^2$ and
electric charge $Q/R_H$. For non-spinning black holes, $J=0$ and
$e=Q^2/R_{H}^{2}$.  The extremal Kerr--Newman solutions which satisfy
$(Q/R_H)^4 + 4 (J/R_H^2)^2 = 1$ all have an extremality parameter
$e=1$.}
\label{KN:QJ}
\end{figure}

Moving on to the Kerr--Newman solutions the functional form of $e$
becomes considerably more complicated and so instead of stating it
explicitly, we plot it in Fig.~\ref{KN:QJ} as a function of the angular
momentum and electric charge.  Recall from Eq.~(\ref{eq:new_kerr_bound})
that the standard extremality condition for Kerr--Newman solutions can
be written as $Q^{4} + 4 J^{2} = R_{H}^{4}$. Then from the figure, we
see that for these extremal solutions $e = 1$ while for non-extremal
solutions $e < 1$.  Therefore, the local extremality quantity behaves as
expected for stationary, asymptotically flat black holes.

\subsection{Trapped surfaces}
\label{Sect:Trapped}

Finally we consider the third characterization of extremality: a horizon
is sub-extremal if there are trapped surfaces ``just inside" the horizon
and extremal (or super-extremal) if there are no such surfaces.  For
this notion we need to understand how the properties of a
$\theta_{(\ell)} = 0$ surface change under infinitesimal deformations.
Continuing to assume that $\theta_{(n)}  < 0$ on the horizon (and so by
continuity remains negative for sufficiently small deformations), we are
then interested in cases where there exists an inward deformation so
that $\theta_{(\ell)}$ also becomes negative. The equations governing
such deformations have been derived and rederived many times and in many
ways over the years \cite{0264-9381-4-2-011, Gourgoulhon:2005ch,
Gourgoulhon:2006uc, Korzynski:2006bx, Eardley:1997hk, Hayward:1993wb,
Booth:2006bn}.  Here we follow \cite{Booth:2006bn}.

There, it is shown that given a spacelike two-surface $S$ and a
transverse deformation vector field $X^a$,
\bea
\delta_X \theta_{(\ell)} = -\tilde{d}^2 B + 2 \tilde{\omega}^a \tilde{d}_a B 
- B \delta_n \theta_{(\ell)} 
+ A \delta_\ell \theta_{(\ell)} \, , 
\label{dXtL}
\eea
where $A$ and $B$ are the components of deformation vector relative to
the null normals so that $X^a = A \ell^a - B n^a$. Furthermore,  
\bea
\delta_\ell \theta_{(\ell)} =
- || \sigma^{(\ell)} ||^{2} - (8 \pi G) T_{ab} \ell^a \ell^b 
\label{dLtL}
\eea
(the Raychaudhuri equation with $\theta_{(\ell)} = 0$) and
\bea
\delta_n \theta_{(\ell)} = - \tilde{R}/2 + || \tilde{\omega} ||^2 
- \tilde{d}_a \tilde{\omega}^a 
+ (8 \pi G) T_{ab} \ell^a n^b \, , 
\label{dntl}
\eea
are the variations of $\theta_{(\ell)}$ under the deformations generated
by the null vectors.  One way of calculating these deformations is to
construct a coordinate system on the manifold for which $S$ is
parameterized by two-coordinates (say $(\theta,\phi)$) and is a level
surface with respect to the other two. The deforming vector (be it $X$,
$\ell$ or $n$) should be a tangent vector field to one of these other
coordinates. For such a construction, quantities such as
$\tilde{\omega}_a$ can be defined on level surfaces of the
non-$(\theta,\phi)$ coordinates in a neighbourhood of $S$. Then the
$\delta$s are Lie derivatives.  In fact we have already seen an example
of this type of construction in this paper; equation
(\ref{eq:lie_l_theta_n}) could equally well be written as $\delta_\ell
\theta_{(n)}$. Indeed, Eq.~(\ref{dntl}) can be obtained from
(\ref{eq:lie_l_theta_n}) by simply switching $\ell$ and $n$ and noting
that $\theta_{(\ell)} = 0$.

Inward deformation vector fields will necessarily take the form $r = A
\ell + B n$ where $B > 0$.  For simplicity, we can rescale the
null vectors such that $B = 1$.  Let us denote these rescaled vectors
as $\bar{\ell}$, $\bar{n}$.  Then 
\bea 
\delta_r \theta_{(\ell)} = \delta_{\bar{n}} \theta_{(\bar{\ell})} \, , 
\eea
since on an isolated horizon $\delta_{\bar{\ell}} \theta_{(\bar{\ell})}
= 0$ regardless of the scaling on the null vectors (note that 
this result is independent of Eq.~(\ref{eq:ih_extrinsic}) ). Thus, slices of
the horizon may be deformed inwards into fully trapped surfaces if and
only if there is a scaling of the null vectors such that
$\delta_{\bar{n}} \theta_{(\bar{\ell})} < 0$.

By this measure we characterize an isolated horizon as sub-extremal if
there exists a scaling of the null vectors such that $\delta_{\bar{n}}
\theta_{(\bar{\ell})} < 0$, extremal if there exists a scaling such
that $\delta_{\bar{n}} \theta_{(\bar{\ell})} = 0$, and
super-extremal if there exists a scaling such that $\delta_{\bar{n}}
\theta_{(\bar{\ell})} > 0$. It is important to keep in mind that each
of these conditions must hold \emph{everywhere} on $\Delta$ and that
it will usually be non-trivial to find the correct scaling needed for
the classification. 

The Kerr solutions themselves provide an example of these difficulties.
For rapidly rotating Kerr black holes with the usual (Killing vector)
scaling of the null vectors, $\delta_n \theta_{(\ell)}$ varies in sign
over $S_v$. However, this does not indicate that these horizons lie
outside the classification system. Instead it suggests that a different
scaling of the null vectors is needed. Such a rescaling is considered
explicitly in Appendix C of \cite{Booth:2006bn}, however here we
generalize that calculation to prove a more general result: for
axi-symmetric isolated horizons with $\theta_{(n)} < 0$ the surface
gravity and trapped surface classifications of extremality are
equivalent. That is
\bea
\mbox{Sub-extremal:} & \kappa > 0 \Leftrightarrow e < 1 \Leftrightarrow 
 \delta_{\bar{n}} \theta_{(\bar{\ell})} < 0 & \, ,  \nonumber \\
\mbox{Extremal:} & \kappa = 0 \Leftrightarrow e = 1 \Leftrightarrow 
\delta_{\bar{n}} \theta_{(\bar{\ell})} = 0 & \, \mbox{and} \nonumber\\ 
\mbox{Super-extremal:} & \kappa < 0 \Leftrightarrow e>1 \Leftrightarrow 
\delta_{\bar{n}} \theta_{(\bar{\ell})} > 0& \, , \nonumber 
\eea
where $\bar{\ell}^a$ and $\bar{n}^a$ are appropriate rescalings of
the null vectors.  

We prove this result by explicitly constructing these rescalings. To
this end we first note the following key fact: On a topologically
spherical two-surface embedded in spacetime there is always a scaling of
the null vectors so that the angular momentum one-form is
divergence-free: $\bar{d}_a \tilde{\omega}^a = 0$ (this result
ultimately follows from the Hodge decomposition theorem
\cite{Ashtekar:2001jb}).  The scaling is unique up to the usual
multiplicative constant. Thus, taking such a pair of vectors
($\ell_o^a$,$n_o^a$) as a reference, we can write any other scaling of
the null vectors as
\bea
\ell^a = f \ell_o^a \; \; \mbox{and} \; \; n^a = \frac{1}{f} n^a_o \, , 
\eea
for some scalar function $f$ over $S_v$. Then we can define the inverse
scaling 
\bea
\bar{\ell}^a = \frac{1}{f} \ell_o^a \; \; \mbox{and} 
\; \; \bar{n}^a = f n^a_o \, 
\eea
and  using equations (\ref{eq:lie_l_theta_n}) and (\ref{dntl}) it is
straightforward to see that on an isolated horizon:
\bea
\delta_{\bar{n}} \theta_{(\bar{\ell})} = 
\kappa \theta_{(n)} - 4 \tilde{\omega}_o^a \bar{d}_a \ln f \, . 
\label{dntl_kappa}
\eea
For an axisymmetric horizon the dual requirements that $\tilde{\omega}_o^a$
respect the symmetry and that $|| \tilde{\omega}_o ||$ not diverge at
the poles of rotation imply that $\tilde{\omega}_o^a$ must be parallel
to the rotation vector. Then the last term of (\ref{dntl_kappa}) 
vanishes and we find that
\bea 
\delta_{\bar{n}} \theta_{(\bar{\ell})} = \kappa \theta_{(n)} \, .
\eea
The result is established. As a corollary, the value of the extremality
parameter $e$ is the same evaluated for either the constant surface gravity 
scaling of the null vectors or the corresponding inverse scaling. 

In more general situations, deciding on the classification of a horizon
will amount to studying the properties of a second order elliptic
partial differential operator. In particular, again taking $(\ell_o,
n_o)$ as reference scalings we have 
\bea
\delta_n \theta_{(\ell)} =  - \tilde{d}^2 \ln f 
+ 2 \tilde{\omega}^a_o \tilde{d}_a \ln f + ||\tilde{d} \ln f||^2 
+ \delta_{n_o} \theta_{(\ell_o)}  
\label{dntl_op}
\eea
where
\bea
\delta_{n_o} \theta_{(\ell_o)} =  || \tilde{\omega}_o ||^2  
- \tilde{R}/2 + 8 \pi G T_{ab} \ell^a_o n^b_o \, , 
\eea
and we wish to find functions $f$ for which (\ref{dntl_op}) is
everywhere negative (or zero or positive). 

We will not investigate this equation in detail in this paper but
instead content ourselves with proving that, in general, the
classification is well defined. That is, given a scaling of the null
vectors so that $\delta_n \theta_{(\ell)}$ is everywhere negative, it is
impossible to rescale the vectors so that the horizon becomes extremal
or super-extremal. Physically this is equivalent to saying that there
cannot be both trapped and untrapped surfaces ``just inside" the
horizon. 

To see this, we reuse Eq.~(\ref{dntl_op}) though this time take
$(\ell_o,n_o)$ as any scaling of the null vectors and $(\ell,n)$ as some
rescaling by $f$. Let us assume for a moment that $f$ is analytic.  Then
if it is not constant it must have maximum and a minimum. At each of
these $\tilde{d} \ln f$ vanishes while $- \tilde{d}^2 \ln f$ is
respectively greater or less than zero. Thus at $f_{max}$: $\delta_{n}
\theta_{(\ell)}  > \delta_{n_{o}} \theta_{(\ell_{o})}$ while at
$f_{min}$: $\delta_{n} \theta_{(\ell)}  < \delta_{n_{o}}
\theta_{(\ell_{o})}$ and it is clear that if $\delta_{n_o}
\theta_{(\ell_o)} = 0$ then $\delta_{n} \theta_{(\ell)} $ will have a
mixed sign.  Similarly if $\delta_{n_o} \theta_{(\ell_o)}$ is everywhere
positive (or negative) then it cannot be rescaled to be everywhere
negative (or positive). These results extend to non-analytic rescalings
with the help of the maximum principle (see for example
\cite{Booth:2006bn}) and so it is clear that rescalings cannot change
the classification of a horizon. 

Finally, note that we have not eliminated the possibility that some
isolated horizons might exist which do not fall into any of the three
categories. That is, there may be horizons for which no rescaling will
cause $\delta_n \theta_{(\ell)}$ to either vanish or be
positive/negative everywhere. A source of potential examples are the
distorted horizons discussed in \cite{Fairhurst:2000xh}.

\subsection{A bound on angular momentum}
\label{Sect:bound}

With these two notions of extremality established let us now return to
the first: is there a maximum angular momentum for rotationally
symmetric isolated horizons?  In answer to this question we now show
that the allowed angular momentum is bound by the intrinsic geometry of
the horizon and in particular for a large class of horizons, that bound
is exactly the local version of the Kerr bound,
Eq.~(\ref{eq:new_kerr_bound}).   For simplicity, in this section, we
restrict our attention to uncharged black holes, i.e. horizons for which
the second term in (\ref{eq:angular_momentum}) vanishes.

First, applying the Cauchy-Schwarz inequality to the definition of
$J[\phi]$, we have 
\begin{eqnarray}\label{eq:ang_mom_bound}
  J[\phi]^{2} &\le& e \times \frac{1}{16 \pi} \int_{S} 
 d^2 x \sqrt{\tilde{q}} ||\phi||^{2}  \, ,
\end{eqnarray}
and so with $e \leq 1$ we immediately have a bound on the angular momentum
determined by the intrinsic geometry of the horizon two-surfaces. 

To better understand this bound we rewrite it as
\bea
J[\phi]^2 \leq e \, \gamma \, R_H^4 \, , 
\eea
where 
\bea
\gamma =  \frac{\int_{S}  d^2 x \sqrt{\tilde{q}} 
||\phi||^{2}}{16 \pi R_H^4} \, . 
\label{gamma}
\eea
The properties of $\gamma$ are more easily studied by introducing a
canonical coordinate system on $S$.  First use the symmetry vector
$\phi^a$ to generate a foliation of $S$ into circles (plus two poles).
Next, choose a point on one of the circles and construct a perpendicular
geodesic from that point. This curve runs from pole to pole and also
perpendicularly intersects each of the other circles. Then the first
coordinate $s$ labels the slices by the proper distance measured along
this geodesic from one of the poles.  Thus $0 \leq s \leq L$ where $L$
is the distance between the poles.  The second coordinate is the usual
rotational coordinate $\phi$ defined so that the geodesic is a curve of
constant $\phi$ and $\phi^a \nabla_a \phi =1$.  For this system the
metric takes the form
\bea
dS^2 = ds^2 + \rho(s)^2 d \phi^2 \, ,   
\eea
where $2 \pi \rho(s)$ is the circumference of the circle of coordinate
radius $s$ and $\rho(0) = \rho(L) = 0$. 

Then (\ref{gamma}) becomes
\bea
\gamma
=  \frac{\int_0^L  \rho^3 d s}{2 \left( \int_0^L  \rho d s \right)^2} \, ,
\eea
and it is clear that for arbitrary $\rho$ there is no bound on $\gamma$
--- given a horizon whose geometry is described by the function $\rho$
and a constant $k$ we can define a new horizon geometry described by
$\rho' = k \rho$ for which $\gamma' = k \gamma$. Thus, $\gamma$ can be
made arbitrarily large. 

However this class of rescalings is only possible if $d \rho / ds $ is
also allowed to become arbitrarily large and, at least in some
circumstances, it is reasonable to bound this quantity. For example, if
$S$ can be embedded in Euclidean $\mathbb{R}^3$ then $d \rho / ds \leq
1$; the maximum rate at  which the circumferential radius can increase
is $d \rho / ds = 1$ (for a flat disc). 

Given a bound  $|d \rho / ds | \leq m$ for some positive constant $m$,
it can be shown that the maximum value of $\gamma$ arises for the curve
that increases with slope $m$ as $s$ runs from $0$ to $L/2$ and then
decreases with slope $-m$ from $L/2$ to $L$ 
(see appendix \ref{MaxGamma}).
In this case
$\gamma = m / 4$ and so (\ref{eq:ang_mom_bound}) becomes
\bea
\frac{J[\phi]^2}{R_H^4} \leq  \frac{m}{4}  \, .  
\label{BoundBound}
\eea

With $m=1$, the angular momentum is bounded by the standard (Kerr) value
(Eq.~\ref{eq:new_kerr_bound}).  Unfortunately not all surfaces of
interest satisfy $|d \rho / ds | \leq 1$. For example it is well known
that sufficiently rapidly rotating Kerr horizons cannot be embedded in
Euclidean $\mathbb{R}^3$.  This is precisely because in this situation
$| d \rho / d s | > 1$ near the poles.  For extremal Kerr it achieves a
maximum value of $3 \sqrt{3}/4 \approx 1.299$ and $\gamma = \pi/4 - 1/2
\approx 0.285$.  Therefore, the local extremality condition
(\ref{eq:extremality}) cannot be used to infer $J \leq R_H^{2}/2$ even
though this condition still holds.  Interestingly, as shown in Appendix
\ref{KadS}, for the Kerr-AdS black holes $J \leq R_H^2/2$ may be
violated by an arbitrary amount.  In such cases $e \le 1$ and the new
bound (\ref{BoundBound}) still holds although with $m = \left| d \rho /
ds \right|_{max}>1$. 

In summary the allowed angular momentum for a rotationally symmetric
isolated horizon is bound by the intrinsic geometry of the surface. For
surfaces that can be embedded in Euclidean $\mathbb{R}^3$ this is
exactly the usual Kerr bound. However for more exotic horizon
cross-sections the intrinsic geometry can be similarly exotic and so the
numerical factor $m$ in (\ref{BoundBound}) can become arbitrarily large.

\section{Dynamical horizons}
\label{sec:dynamical}
 
Let us now turn our attention to local, interacting horizons.  There are
several formulations describing these horizons, but here we choose to
use the dynamical horizon framework of Ashtekar and Krishnan
\cite{Ashtekar:2002ag, Ashtekar:2003hk}. The results are equally
applicable, with minor modifications, to the other formulations,
including Hayward's trapping horizons \cite{Hayward:2004fz,
Hayward:2006he, Hayward:2004dv}.

We begin by recalling a few basic properties of dynamical horizons.  A
dynamical horizon $H$ is a spacelike three-surface, uniquely
\cite{Ashtekar:2005ez}  foliated by two-surfaces which are marginally
trapped, namely $\theta_{(\ell)} = 0$ and $\theta_{(n)} < 0$. Given a
foliation label $v$ we can write the evolution vector field
$\mathcal{V}^a$, defined so that it is normal to the two-surfaces and
$\Lie_{\mathcal{V}} v = 1$, as 
\begin{equation}
  \mathcal{V}^{a} = A \ell^{a} - B n^{a} 
  \label{cV}
\end{equation}
for some functions $A$ and $B$. Since $\mathcal{V}^a$ is tangent to the
spacelike $H$, neither $A$ nor $B$ can vanish anywhere.  Further if we
choose the orientation of the labelling so that $A>0$ it follows that
$B>0$ as well.  From this and the assumption that $\theta_{(n)} < 0$, it
immediately follows that, like event horizons, dynamical horizons always
expand in area:
\bea
\mathcal{L}_{\mathcal{V}} \sqrt{\tilde{q}} = 
- \sqrt{\tilde{q}} B \theta_{(n)} \, . 
\eea

We now consider the possible characterizations of extremality. The
ambiguities in defining a mass or energy are even greater than for
isolated horizons so we do not pursue that characterization. There are
also difficulties in using surface gravity. While this can be defined 
analogously to the surface gravity on an isolated horizon:
\begin{equation}\label{eq:dh_kappa}
  \kappa = - \mathcal{V}^{a} n_{b} \nabla_{a} \ell^{b} \, ,
\end{equation}
it is generally not constant on a dynamical horizon.  This is to be
expected as, taking the analogy between black holes and thermodynamics,
it is equivalent to the statement that the temperature will not normally
be constant for a system away from equilibrium. Thus, for dynamical
horizons we have little control over the value of the surface
gravity, and hence cannot use it in constructing an extremality
condition.

A partial exception to this statement occurs if the horizon is slowly
evolving as defined in \cite{Booth:2003ji,Kavanagh:2006qe,Booth:2006bn}.
In this case it is in quasi-equilibrium and ``almost" isolated and one
can show that the surface gravity almost constant. It changes slowly
both across the two-surface cross sections and in evolving up the
horizon.  In this case it would be feasible to state that the horizon is
non-extremal, at least to the order for which $\kappa$ is constant.

More generally we are left to consider the implication of the existence
of fully trapped surfaces just inside the horizon.  Here, we restrict
attention to ``generic dynamical horizons'' in the sense of
\cite{Ashtekar:2005ez}, in order that we are dealing only with black
hole horizons and not cosmological horizons or horizons arising in other
space-times, such as those in \cite{Senovilla:2003tw}.  The genericity
condition requires that $\delta_\ell \theta_{(\ell)}$ does not vanish at
any point on the horizon.  As has been shown previously, for example in
Refs.~\cite{Hayward:1993wb,Ashtekar:2003hk}, this guarantees that there
will be trapped surfaces inside the horizon.   Here we briefly repeat
the argument.

First, scale the null vectors so that
\begin{equation}\label{eq:dh_ell}
  \underleftarrow{\ell} \propto dv \, ,
\end{equation}
or equivalently 
\bea
\ell^a = f(v) (\hat{r}^a + \hat{\tau}^a) \, ,
\label{eq:DHell} 
\eea
where $f(v)$ is a positive function, $\hat{\tau}^a$ is the
future-pointing timelike normal to the horizon and $\hat{r}^a$ is the
in-horizon spacelike normal to the slices that points in the direction
of increasing area.  For such a scaling, $B$ in Eq.~(\ref{cV}) is

constant on cross sections of the horizon whence Eq.~(\ref{dXtL})
simplifies to 
\bea
\Lie_{\mathcal{V}} \theta_{(\ell)} =
A \delta_\ell \theta_{(\ell)} - B \delta_n \theta_{(\ell)} \, . 
\eea 
Thus, with $\theta_{(\ell)}$ zero everywhere on $H$, 
\bea
\delta_n \theta_{(\ell)} = 
\left( \frac{A}{B} \right) \delta_\ell \theta_{(\ell)} \, . 
\eea
Finally if the null energy condition holds and we assume that
$\delta_{\ell} \theta_{(\ell)}$ nowhere vanishes, then by
Eq.~(\ref{dLtL}) we have
\begin{equation}\label{eq:dh_delta_n_theta_l}
\delta_n \theta_{(\ell)} < 0 \, . 
\end{equation}

Since trapped surfaces must exist inside a dynamical horizon, we can
immediately apply the results of sections \ref{Sect:Trapped} and
\ref{Sect:bound} to them.  In particular, it follows that the
extremality parameter $e$ introduced in Eq.~(\ref{eq:extremality})
cannot exceed unity.  Further, by
Eq.~(\ref{eq:dh_delta_n_theta_l}) it must be strictly less than one
--- dynamical horizons must be sub-extremal.
Thus, for example, in Einstein-Maxwell theory the angular momentum
one-form and electric and magnetic fluxes are bound on a dynamical
horizon by 
\begin{equation}
  \frac{1}{4 \pi} \int_{S_v} d^2 x \sqrt{\tilde{q}} 
   \left( ||\tilde{\omega}||^{2}  + G (E_\perp^2 + B_\perp^2) \right) 
  < 1 \, ,
\end{equation}
and for an axi-symmetric horizon whose cross-sections can be embedded
in Euclidean $\mathbb{R}^3$:
\bea
J[\phi]^2 < R_H^4/4 \, . 
\eea

\begin{figure}
\includegraphics{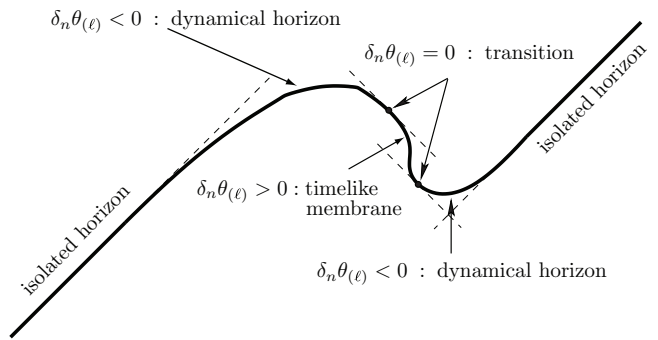}
\caption{A schematic of a horizon ``jump". Matter falls into an isolated
horizon causing it to expand as a dynamical horizon. However at a
certain point the density of the matter is such that a new horizon forms
outside the old resulting in a jump which geometrically corresponds to a
\emph{timelike membrane} (a timelike three-surface with $\theta_{(\ell)} = 0$,
$\theta_{(n)} < 0$) connecting two
dynamical horizons. In this figure $45^\circ$ lines are null, time
increases in the vertical direction, and surface area increases to the
right.}
\label{MTT_fig}
\end{figure}

While dynamical horizons cannot violate the trapped surface extremality
condition, they can transform into a new type of structure that does not
contain trapped surfaces inside. In \cite{Booth:2005ng}, it is
demonstrated that marginally trapped tubes (foliated three-surfaces
which satisfy $\theta_{(\ell)} = 0$ and $\theta_{(n)} < 0$) in
Tolman-Bondi spacetimes may transform from being spacelike (and so
dynamical horizons with $\delta_n \theta_{(\ell)} < 0$) to timelike
(with $\delta_n \theta_{(\ell)} > 0$).  The change in behaviour occurs
when the dust density $\rho$ becomes greater than $1/a$ where $a$ is the
area of the horizon cross-sections. Intuitively one can think of this as
occurring when the matter density becomes high enough to form a new
horizon outside the old and then the timelike section of the horizon is
characteristic of a horizon ``jump". Such a behaviour is shown in Fig.
\ref{MTT_fig} (which is adapted from \cite{Booth:2005ng}). 

Similarly, Schnetter, Krishnan, and Florian  \cite{Schnetter:2006yt}
have studied various numerical simulations including the collision of
spinning black holes.  They find that before the holes collide, an outer
spacelike horizon forms.  At the same time, an inner horizon forms which
is part spacelike, part timelike.  Although they were unable to follow
the evolution far enough, they conjecture that the horizons will form a
continuous three-surface, only some fraction of which is spacelike.  The
results presented here suggest a criterion for determining when these
jumps are about to  occur --- the transition between the spacelike and
timelike sections of the horizon will occur as $e \rightarrow 1$. Note
however that in general this transition may be complicated and include
horizon cross-sections whose evolution may be spacelike in some areas
and null or timelike in others. For a detailed understanding of the
transition one would need to track $\delta_n \theta_{(\ell)}$ and/or the
signature of the evolution vector point-by-point.  However while the
evolution is still purely spacelike, $e$ should provide a good estimate
of the proximity to extremality. 

With a view towards tracking either $e$ or $\delta_n \theta_{(\ell)}$ in
a simulation, let us reformulate the expressions above in terms of
spacelike/timelike unit normals.  First, in terms of the
unit tangent ($\hat{r}^a$) and normal ($\hat{\tau}^{a}$) vectors to the
horizon, the angular momentum one-form and stress-energy component can
be rewritten as 
\bea
\tilde{\omega}_a = \tilde{q}_a^b \hat{r}^c K^{(\hat{\tau})}_{bc}
 \; \; \mbox{and}
\; \; T_{ab} \ell^a n^b = 
T_{ab} (\hat{\tau}^a \hat{\tau}^b -  \hat{r}^a \hat{r}^b) 
\eea
for the preferred scaling (\ref{eq:DHell}) where $K^{(\hat{\tau})}_{bc}$
is the extrinsic curvature of the horizon relative to $\hat{\tau}$.

Alternatively we can consider an apparent horizon, with unit normal
$\hat{s}$, in a three-slice $\Sigma_t$, with unit timelike normal
$\hat{u}$.  The horizon evolution vector field $\mathcal{V}$ can be
expressed as
\begin{equation}
  \mathcal{V}^{a} = N (\hat{u}^{a} + v_{\perp} \hat{s}^{a})
\end{equation}
where $N$ is the lapse and $v_{\perp}$ is the the velocity of the
horizon relative to the foliation.  Then, following \cite{Booth:2007ix},
we can write
\bea
\delta_{n} \theta_{(\ell)} 
= - \frac{1}{2} \left( \frac{v_\perp + 1}{v_\perp - 1} \right)
(  || \sigma_{(\underline{\ell})} ||^2 + 8 \pi G T_{ab} \underline{\ell}^a 
\underline{\ell}^b)  
\label{Extremal}
\eea
where $\underline{\ell}^a = \hat{u}^a + \hat{s}^a$ and so
$\sigma_{(\underline{\ell}) ab} = \tilde{q}_a^c \tilde{q}_b^d (K_{cd} + D_c
\hat{s}_d )$.  We obtain a similar expression for the extremality parameter $e$
as:
\bea
  e &=& 1 +\frac{1}{4 \pi G} \int_{S_t} \mspace{-10mu} d^2 x 
    \sqrt{\tilde{q}} \delta_{n} \theta_{(\ell)} \\ 
  &=& 1 - \frac{1}{8 \pi G} \int_{S_t}  \mspace{-10mu} d^2 x
    \sqrt{\tilde{q}} \left( \frac{v_\perp + 1}{v_\perp - 1} \right)
    (  || \sigma_{(\underline{\ell})} ||^2 + 8 \pi G 
    T_{ab} \underline{\ell}^a \underline{\ell}^b)  
    \nonumber \, . 
\label{e}
\eea
See Ref.~\cite{Booth:2007ix} for further details of these calculations. 

Thus, thanks to the preferred scaling of the null normals we have an
unambiguous definition of $e$ on each slice of any dynamical horizon. 
Things are, however, slightly more complicated in more general
situations. First, keep in mind that the scalings (\ref{eq:DHell}) are
defined by the timelike normal to $H$ and the spacelike normal to 
$S_v$ in $H$. Thus one cannot use this form of the definition if
$H$ becomes null (either as an isolated horizon or while transitioning
to become a timelike membrane). In such cases one must return
to the original definition Eq.~(\ref{eq:extremality}).  

There is also a second situation where it is not feasible to use 
Eq.~(\ref{e}) to calculate $e$. An apparent horizon $S$ 
in a set of initial data can evolve into many different dynamical
horizons depending on
how the data itself is evolved -- that is apparent horizons are foliation dependent. From the point of view of  Eq.~(\ref{e}) the various 
potential horizons will generate different scalings of the null normals 
to $S$ and so different
values of $e$. This ambiguity can be more easily understood by  switching back to the original definition of the extremality parameter 
given in Eq.~(\ref{eq:extremality}). Then, if $\ell' =  f \ell$ and $n' = n/f$
the ambiguity in $e$ under rescalings of the null vectors is given by 
\bea
e' =   e  + \frac{1}{4\pi} \int_{S} \mspace{-10mu} 
d^2 x \sqrt{\tilde{q}} 
\left( 2 \tilde{\omega}^a \tilde{d}_a \ln  f + || d \ln f||^2  \right) 
\eea
while 
\bea
\delta_{n'} \theta_{(\ell')} = \delta_n \theta_{(\ell)} - \tilde{d}^2 \ln f + 
||d \ln f||^2 + 2 \tilde{\omega}^a \tilde{d}_a \ln f \, .
\label{dntlVar}
\eea
That said, it should be kept in mind that by the trapped surface 
classification presented in Section \ref{Sect:Trapped}, $S$ is
defined as sub-extremal, extremal, or super-extremal based on
possible rather than any particular scalings of the null vectors. 
Though $e$ may vary for various choices of scalings, 
the ultimate classification of $S$ is invariant. 
For dynamical horizons a suitable scaling is defined by 
the normals to $H$, but if one only has a single surface, then 
one must go back to an analysis of 
the elliptic operator defined by Eq.~(\ref{dntlVar}).

Interestingly, in the context of trapping horizons, Hayward
\cite{Hayward:1993wb} has introduced an alternative expression for
surface gravity which is proportional to $\sqrt{-\delta_{n}
\theta_{(\ell)}}$.  Such a definition explicitly ties together the
non-vanishing of surface gravity with the existence of trapped surfaces
inside the horizon, i.e. the second and third characterizations of
extremality necessarily coincide.  Furthermore, making use of
Eq.~(\ref{eq:lie_l_theta_n}) he has obtained a zeroth law for trapping
horizons which has many similarities with the extremality condition
introduced in this paper.

\section{Summary}
\label{sec:summary}

In this paper, we have considered three characterizations of
extremality. The first is the standard Kerr bound on angular momentum
relative to mass. We have argued that in general it is not well-posed
due to the difficulties in defining mass and angular momentum in general
relativity.  Even when the Kerr bound is reformulated in terms of
horizon area and angular momentum, it can only be meaningfully evaluated
on axi-symmetric horizons.  Furthermore, while we have not provided an
explicit violation of this bound in asymptotically flat spacetimes, we
have argued that it is likely that it can be violated.  In particular
Kerr-AdS solutions can violate the bound by an arbitrary amount. These
results do not violate the recent theorems of Dain which only apply to
asymptotically flat vacuum spacetimes.  

A more satisfactory characterization of extremality for isolated
horizons arises from the surface gravity.  For isolated horizons, a
sub-extremal horizon will have positive surface gravity, while the
surface gravity for an extremal horizon vanishes.  Furthermore,
non-negativity of surface gravity leads to a bound on the integrated
square of the angular momentum density and the matter stress-energy at
the horizon. 

Alternatively, we can characterize non-extremality as the requirement
that there should be fully trapped surfaces just inside a black hole
horizon. This notion is then applicable to both isolated and dynamical
horizons.  In addition, this condition again leads to a bound on the
integrated square of the angular momentum density and the matter
stress-energy at the horizon.  The surface gravity and trapped surface
characterizations of extremality for isolated horizons are very closely
related, and indeed in axi-symmetry are entirely equivalent.

The local extremality condition is also sufficient to place a
restriction on the maximum allowed angular momentum relative to the
intrinsic geometry of the horizon.  For horizons whose cross-sections
can be embedded in Euclidean $\mathbb{R}^3$ this is sufficient to imply
the standard Kerr bound $4J^2 < R_H^4$, however for more exotic intrinsic
geometries we can only show that $4J^2 < m R_H^4$ for some constant $m$
which may be made arbitrarily large in, for example, Kerr-AdS. 

Thus, the notion of extremality extends beyond the Kerr solutions though
in a non-trivial way.  The spirit of the bounds remains. The angular
momentum of the horizon is bounded relative to the intrinsic geometry of
the horizon.  In general, when no axis of rotation exists, it is the
square of the angular momentum density which is bounded.  Equivalently,
one can think of this as a bound on the sum of the multipole moments of
the angular momentum rather than the dipole itself.  The extremality
quantity, $e$, which we have introduced should be calculable on apparent
horizons occuring in numerical relativity simulations and would provide
an interesting characterization of how close to extremality a black hole
is immediately following a merger.  

\section*{Acknowledgements}

We would like to thank Peter Booth, Patrick Brady, Jolien Creighton,
Herb Gaskill and Badri Krishnan for helpful discussions.  Ivan Booth was
supported by the Natural Sciences and Engineering Research Council of
Canada.  Stephen Fairhurst was supported by NSF grant PHY-0200852 and
the Royal Society.

\appendix

\section{Kerr anti-deSitter black holes}
\label{KadS}


\begin{figure}
\includegraphics{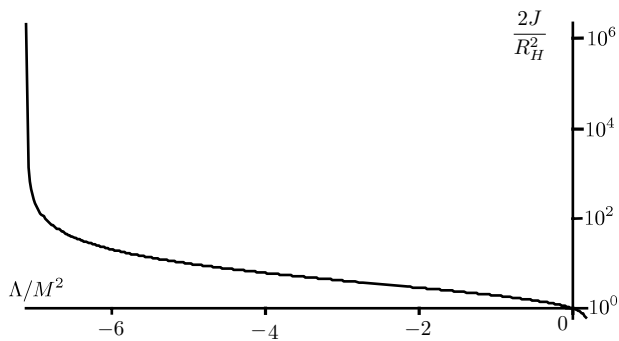}
\caption{$2J/R_H^2$ versus $\Lambda$ for the extremal Kerr-(A)dS family
of black holes. For $\Lambda=0$, $2J/R_H^2 = 1$, as expected for the
extremal, asymptotically flat Kerr solution.  For $\Lambda > 0$,
$2J/R_H^2 < 1$ while for $\Lambda<0$, $J/R_H^2 > 1/2$ and it diverges as
$\Lambda$ approaches its minimum allowed value.  Therefore, for extremal
Kerr anti-deSitter black holes, the local reformulation of the Kerr
bound (\ref{eq:new_kerr_bound}) is violated.}
\label{KadS_fig}
\end{figure}

The Kerr-(anti)deSitter family of solutions are described by the metric:
\bea
ds^2 & =  & - \frac{\Delta}{\rho^2} 
\left(dt - \frac{a}{\Xi} \sin^2 \theta d \phi \right)^2 
+ \frac{\rho^2}{\Delta} dr^2 + \frac{\rho^2}{\Delta_\theta} 
d \theta^2 \nonumber \\
& & + \frac{\Delta_\theta \sin^2 \theta}{\rho^2} 
\left(a dt - \frac{r^2+a^2}{\Xi} d \phi \right)^2 
\label{eq:kads}
\eea
where
\bea
\Delta & = & - \frac{\Lambda}{3} r^4 + \left(1-\frac{\Lambda}{3}a^2 \right) r^2 - 2 M r + a^2 \, , \nonumber \\
\Delta_\theta & = & 1 + \frac{\Lambda}{3} a^2 \cos \theta^2  \, , \nonumber\\
\Xi & = & 1 + \frac{\Lambda}{3} a^2 \, \mbox{and} \nonumber\\
\rho^2 & = & r^2 + a^2 \cos^2 \theta \, . 
\eea
$\Lambda$ is the cosmological constant (and so is positive for deSitter
and negative for anti-deSitter), $M$ is the mass parameter, and $a$ is
the rotation parameter. 

We are interested in the black hole sector of the solution space. The
various horizons occur at the roots of $\Delta = 0 $.  For $\Lambda>0$,
$\Delta$ has four roots in the black hole sector. In increasing order
they are a (negative) unphysical solution, inner black hole horizon,
outer black hole horizon, and the cosmological horizon.  For $\Lambda<0$
there are just two roots: the inner and outer black hole horizons. Our
interest is in the outer black hole horizon which we label $r_+$. The
coordinate representation (\ref{eq:kads}) of the metric diverges at
$r_+$ but for our purposes we can work around this by considering
appropriate limiting cases which are well-defined.  Then, one can show
(see for example \cite{Dehghani:2001af}) that the areal radius and
angular momentum of the horizon are, respectively, 
\bea
R_H^2 = \frac{r_+^2 + a^2}{\Xi} \; \; \mbox{and} 
\; \; J[\phi] = \frac{Ma}{\Xi^2} \, . 
\eea

Here, for definiteness, we focus on the extremal horizons of this family
where the inner and outer horizons coincide and so $\Delta$ has a
degenerate root (the second and third roots are degenerate for
$\Lambda>0$). Such cases are most easily identified by examining where
the discriminant of $\Delta$ vanishes (the expression is a quintic in
$a^2$ and quartic in $\Lambda$ but may be dealt with easily enough with
the help of a computer algebra system).

For a given value of the mass
$M$, there is a finite range of $\Lambda$ for which extremal solutions
exist. The lower bound is  $\Lambda_{\mathrm{min}} \approx -7.1/ M^{2}$ 
where $\Xi = 0$ (this is a lower bound as for $\Xi < 0$ the signature of the
the $\theta$ coordinate changes and becomes timelike close to $0$ and $\pi$).  
The maximum
value $\Lambda_{\mathrm{max}} \approx 0.18 / M^{2}$ occurs when the
inner and outer black hole horizons and the cosmological horizon all
coincide in a triply degenerate root.

Given the range of values of $\Lambda$ which permit an extremal horizon,
we can plot $J/R_H^2$ and see whether the extremality bound
(\ref{eq:new_kerr_bound}) is violated.  This is shown in Fig.~\ref{KadS_fig} 
and it is clear that the bound $J \leq R_H^2/2$ is
violated for all extremal Kerr-AdS solutions. To understand this in
light of the discussion in Section \ref{Sect:bound}, first note that in
the presence of a cosmological constant Eq.~(\ref{eq:extremality})
becomes:
\begin{equation}\label{eq:extremalityL}
  e := \int_{S_v} d^2 x \sqrt{\tilde{q}} 
   \left( 2G T_{ab} \ell^{a} n^{b} + \frac{1}{4\pi}||\tilde{\omega}||^{2}  
  - \Lambda \right) \le 1 \, ,
\end{equation}

Thus for $\Lambda < 0$, the contribution from the cosmologicl constant
is positive and so we still have $J^2/R_{H}^{4} \leq m/4$. In this case
however $m$ becomes arbitrarily large as we approach $\Xi = 0$.
Specifically, the induced metric on a cross-section of the horizon is
\bea 
dS^2 = \frac{\rho^2}{\Delta_\theta} d\theta^2 + 
\frac{\Delta_\theta (r^2+a^2)^2 \sin^2 \theta}{\rho^2 \Xi^2} d\phi^2 \, .  
\eea
Then the circumferential radius is 
\bea 
R = \frac{\sqrt{\Delta_\theta} (r^2+a^2) \sin \theta}{\rho \Xi} 
\eea 
and
\bea 
m = \mbox{Max} \left( \frac{\sqrt{\Delta_\theta}}{\rho} 
\frac{dR}{d\theta} \right) \, .  
\eea
It is easy to see that this quantity diverges as $\Xi \rightarrow 0$. 

For simplicity we only considered extremal horizons here, but it is
clear (by continuity) that these violations of the bound will also
extend into parts of the non-extremal sector. 

Finally, it is perhaps interesting to note that on inserting a $\Lambda
> 0$ into Eq.~(\ref{eq:extremalityL}), we see that the upper bound on
the integral of $|| \tilde{\omega} ||^2$ increases with increasing
$\Lambda$. However, at least for Kerr-dS this does not provide enough
freedom to violate $J^2 \leq R_H^4$ as there is a concomitant tightening
of $m$.

\section{Surface of maximum $\gamma$}
\label{MaxGamma}

Let $R$ be the set of continuous functions $\rho(s): [0,L] \rightarrow \mathbb{R}$ that satisfy
\renewcommand{\labelenumi}{\roman{enumi})}
\begin{enumerate}
\item $\rho(s) \geq 0$
\item $\rho(0) =  \rho(L) = 0$, 
\item $d \rho / ds (0) > 0$ and $d \rho/ds (L) < 0$ and
\item $|d \rho / ds | \leq m$ for some $m >0$. 
\end{enumerate}
Thinking back to the two-surfaces defined by these $\rho$, the first condition guarantees a non-negative ``radius", the second and third require that the surfaces close exactly at $0$ and $L$, and
the fourth is the assumed bound on the maximum rate of change of the radius relative to 
the arclength. 

\begin{figure}
\includegraphics{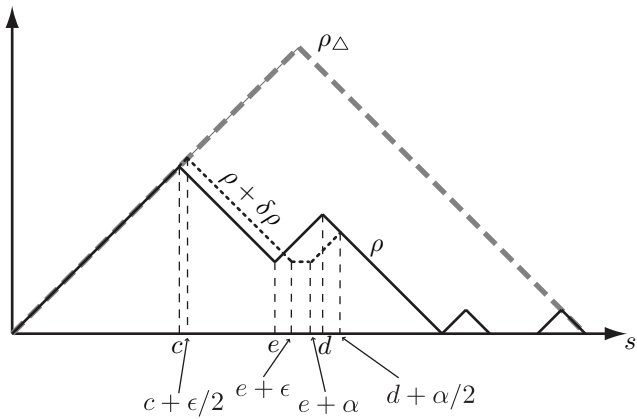}
\caption{Several $\rho$ that appear in the text. $\rho_\triangle$ is the maximizing curve, $\rho$ is a
typical ``saw-toothed" curve, and $\rho + \delta \rho$ (which appears as a dotted line where it 
doesn't coincide with $\rho$) is a variation of that curve which increases the value of $\gamma$. }
\label{sawtooth}
\end{figure}

Further define
$$
\gamma
=  \frac{\int_0^L  \rho^3 d s}{2 \left( \int_0^L  \rho d s \right)^2} . 
$$
Then in this appendix we show that of all $\rho \in R$, the triangular function
\bea
\rho_{\triangle}(s) = \left\{
\begin{array}{ll}
m s & 0 \leq s \leq L/2 \\
m (L - s) & L/2 \leq L 
\end{array}
 \right. \, ,  \label{Max}
\eea
shown in Fig.~\ref{sawtooth} maximizes $\gamma$. 

This is slightly more complicated than a basic variational problem. As noted in the text, if one generalizes to the set of all non-negative functions then $\gamma$ is unbounded. Thus, our 
goal is to show that $\gamma$ is globally maximized over $R$  by the ``boundary" curve 
$\rho_\triangle$. 

To prove this we first show that $\rho_\triangle$ gives a local maximum. To this end, we calculate
 the first variation of $\gamma$ in $\rho$ as 
\bea
\delta \gamma = \frac{\left( 3 \int \mspace{-4mu} \rho^2 \delta \rho ds \right) 
\left(\int \mspace{-4mu} \rho ds \right) - 2 \left( \int \mspace{-4mu} \delta \rho ds \right)}{2 \left(\int \mspace{-4mu} \rho ds \right)^3} \, , \label{firstvar}
\eea
where all integrals are from $0$ to $L$. For variations around $\rho_\triangle$ this becomes
\bea
\delta \gamma_\triangle = \frac{8}{L^4} \int_0^{\frac{L}{2}} (12 s^2 - L^2) 
( \delta \rho (s) +   \delta {\rho} (L - s) ) ds \, . 
\eea 
Now by the restriction on the maximum slope, all allowed $\delta \rho \leq 0$ and further
$\delta \rho$ is non-increasing from $0$ to $L/2$. Thus taking $s_o = L/(2\sqrt{3})$
(the zero of $12 s^2 -L^2$) as a dividing point , we have 
$|\delta \rho(s)|  \leq \delta \rho(s_o)$ for $s \in [0,s_o]$ and 
$|\delta \rho(s)|  \geq \delta \rho(s_o)$ for $s \in [s_o, L/2]$. 
Similar results apply for $\delta {\rho} (L - s)$ which is also non-increasing on this interval. 
Then, keeping in mind that $\delta \rho$ must be zero at least somewhere we find
$\delta \gamma_\triangle < 0$. That is, all allowed variations decrease the value of $\gamma$ 
and so $\rho_\triangle$ provides at least a local maximum for our problem. 

We complete the proof by showing for any other $\rho \in R$ we can find a $\gamma$-increasing
variation $\delta \rho$. It will be sufficient to restrict our attention to the subset of variations 
for which $\int \mspace{-4mu} \delta \rho ds = 0$. For such variations (\ref{firstvar}) simplifies and
we find
\bea
\delta \gamma > 0 \Leftrightarrow \int \mspace{-4mu} \rho^2 \delta \rho ds > 0 \, . 
\label{simpGam}
\eea
Intuitively these inequalities can be satisfied by constructing variations which increase $\rho$ where 
it is larger and balancing this off by decreasing it where it is smaller. 

First consider the case where there is an interval $[a,b]$ over which $\rho$ is monotonically
increasing but $d \rho / ds < m$ and construct a variation
\bea
\delta \rho = \left\{ 
\begin{array}{ll}
0 & 0 \leq s < a \\
- \epsilon \sin \left( 2 \pi \left(  \frac{s-a}{b-a} \right) \right) & a \leq s \leq b\\
0 & b < s \leq b\\
\end{array}
\right.  \, , \label{var1}
\eea
where $\epsilon$ is arbitrarily small;  in particular it is sufficiently
small to ensure that $\rho + \delta \rho > 0$ and $|d (\rho + \delta \rho) / ds | < m$.
By the monotonicity $ \rho(s) < \rho ((a+b)/2) $ for $s \in [a ,(a+b)/2 )$
and $\rho(s) > \rho ((a+b)/2) $ for $s \in ((a+b)/2,b]$,
so by (\ref{simpGam}) it is straightforward to see that $\delta \gamma > 0$.
Thus any $\rho$ that contains an increasing region over which $d \rho / ds < m$, cannot 
maximize $\gamma$. A nearly identical argument shows that an $\rho$ with a decreasing region over which $-m < d \rho / ds$ cannot provide a maximum. 

In fact the same variation (\ref{var1}) can also be used to eliminate all $\rho$ which 
contain a constant section $[a,b]$ over which $\rho = \rho_o>0$. 
In that case $\delta \gamma$
vanishes but a straightforward calculation of the second variation shows that this is because
such a $\rho$ is a local minimum with respect to these variations.

Thus a $\rho$ which maximizes $\gamma$ must have slope $\pm 1$ everywhere -- that 
is either $\rho_\triangle$ or a (possibly broken) ``saw-toothed" curve such as that shown in 
Fig.~\ref{sawtooth}. 
There are several special cases to consider here but while details differ, the basic variation
is the same: we increase a higher peak while decreasing a lower one and so increase $\gamma$. 
In the interests of saving space we consider only the case of two immediately adjoining peaks as
shown in the figure. 

Then, with the higher peak at $c$, lower at $d$ and the valley in between at $e$ we consider
variations of the following type:
\bea
\delta \rho = \left\{
\begin{array}{ll}
2 m (s-c) & c \leq s \leq c + \epsilon/2 \\
m \epsilon & c + \epsilon /2 \leq e\\
m\epsilon - 2 m (s-e) & e \leq s \leq e  + \epsilon \\
- m (s - e) & e + \epsilon \leq s \leq e + \alpha \\
- m \alpha & e + \alpha \leq s \leq d\\
- m \alpha + 2 m (s-d) & d \leq s \leq d + \alpha/2 \\
\end{array} 
\right. \; . 
\eea
$\epsilon$ is the usual small parameter and $\alpha$ is chosen so that $\int \delta \rho ds = 0$. To 
first order (which is all that is needed for a variational calculation) it is
\bea
\alpha \approx \left(\frac{e-c}{d-e}\right) \epsilon \, .  
\eea
Then a direct calculation with (\ref{simpGam}) shows that if the first peak is higher than the second, 
$\delta \gamma > 0$. If they are equal then $\delta \gamma = 0$ but going to the second order
variation, it can be seen that this is because it is a local minimum under such variations. 
Similarly ponderous calculations can be performed to show that no 
other ``saw-toothed" $\rho$ is a
maximum. 

Thus in summary we have shown that $\rho_\triangle$ is a local maximum for 
curves in $R$ while there exist variations of all other curves that increase $\gamma$. Thus, 
$\rho_\triangle$ is the global maximum as claimed.

\bibliography{bhpapers}

\end{document}